\documentstyle[11pt,newpasp,twoside,epsfig]{article} 
\newcommand{\Lya}{Ly-$\alpha$\thinspace} 
 
\newcommand{\eg}{e.g.\thinspace} 
\newcommand{\ie}{i.e.\thinspace} 
\newcommand{\etal}{et al.\thinspace} 
\def\HI{\mbox{H{\scriptsize I}}}

\def\edcomment#1{\iffalse\marginpar{\raggedright\sl#1\/}\else\relax\fi} 
\reversemarginpar 
 
\begin{document} 
\title{Numerical studies of galaxy formation using special purpose hardware} 
\author{Matthias Steinmetz} 
\affil{Steward Observatory, University of Arizona, 933 N Cherry Ave, Tucson, AZ 
85721, USA} 
 
\begin{abstract} 
I review recent progress in numerically simulating the formation and evolution
of galaxies in hierarchically clustering universes. Special emphasis is given to
results based on high-resolution gas dynamical simulations using the N-body
hardware integrator GRAPE. Applications address the origin of the spin of disk
galaxies, the structure and kinematics of damped \Lya systems, and the origin of
galaxy morphology and of galaxy scaling laws. 
\end{abstract} 
 
\section{Introduction} 
 
Motivated by the increasing body of evidence that most of the mass in the
universe consists of invisible ``dark'' matter, and by the particle physicist's
inference that this dark matter is made of exotic non-baryonic particles, a new
and on the long run more fruitful approach to study the formation of galaxies
has been developed: rather than to model the formation and evolution of galaxies
from properties of present day galaxies, it is attempted to prescribe a set of
reasonable initial conditions. The evolution of galaxies is then modeled
starting from these initial conditions. Physical processes are taken into
account that are considered to be relevant such as gravity, gas dynamics,
radiative cooling and star formation. The outcome at different epochs is then
confronted against observational data.
 
One scenario that has been extensively tested in that way is the model of
hierarchical clustering, currently the most successful paradigm of structure
formation in the universe. In this scenario, structure grows as objects of progressively larger
mass merge and collapse to form newly virialized systems. The probably best know
representative of this class of models is the {\sl Cold Dark Matter} (CDM)
scenario. The initial conditions consist of the cosmological parameters
($\Omega, \Omega_{\rm baryon}, \Lambda, H_0$) and of an initial fluctuation
spectrum such as the CDM spectrum. The remaining free parameter, the amplitude
of these initial fluctuations, is calibrated by observational data, \eg, the
measured anisotropies of the microwave background. Over the past few years,
limits on the values allowed for these parameters have been consistently refined
by improved observational techniques and theoretical insight, and it is widely
accepted that a new ``standard'' model has emerged as the clear front-runner
amongst competing models of structure formation. This $\Lambda$CDM model
envisions an eternally expanding universe with the following properties (Bahcall
\etal 1999): (i) matter makes up at present less than about a third of the
critical density for closure  ($\Omega_0 \approx 0.3$); (ii) a non-zero
cosmological constant restores the flat geometry predicted by most inflationary
models of the early universe ($\Lambda_0=1-\Omega_0\approx 0.7$); (iii) the present
rate of universal expansion is $H_0 \approx 70$ km s$^{-1}$ Mpc$^{-1}$ ($h=H_0/100$
km s$^{-1}$ Mpc$^{-1} \approx 0.7$); (iv) baryons make up a very small fraction of
the mass of the universe ($\Omega_b \approx 0.019 \, h^{-2} \ll \Omega_0$); and
(v) the present-day {\sl rms} mass fluctuations on spheres of radius 8 $h^{-1}$
Mpc is of order unity ($\sigma_8 \approx 0.9$).  The hierarchical structure
formation process in this $\Lambda$CDM scenario is illustrated in Figure 1,
which depicts the growth of structure within a $32.5/h\,$Mpc box between
redshifts nine and zero.  The $\Lambda$CDM model is consistent with an
impressive array of well-established fundamental observations such as the age of
the universe as measured from the oldest stars, the extragalactic distance scale
as measured by distant Cepheids, the primordial abundance of the light elements,
the baryonic mass fraction of galaxy clusters, the amplitude of the Cosmic
Microwave Background fluctuations measured by COBE, BOOMERANG, MAXIMA and DASI,
the present-day abundance of massive galaxy clusters, the shape and amplitude of
galaxy clustering patterns, the magnitude of large-scale coherent motions of
galaxy systems, and the world geometry inferred from observations of distant
type Ia supernovae, among  others. 
 
\begin{figure*} 
\epsfig{file=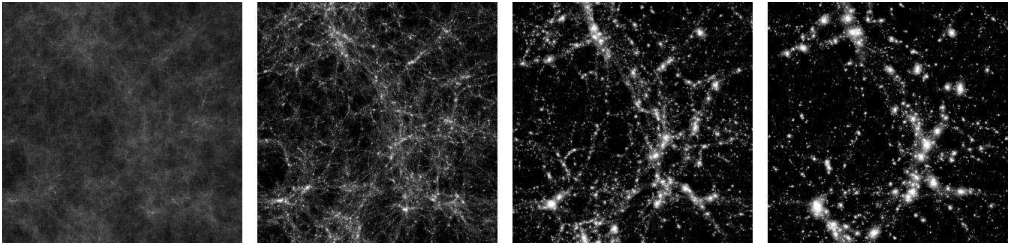,height=3.2cm} 
\caption[]{Time sequence of structure formation in a hierarchical clustering universe,  
here for the so-called $\Lambda$CDM model. The four snapshots correspond (from left to 
right) to redshifts of 9, 3.5, 1 and 0, respectively. The side length of the 
simulation box is $32.5\,h^{-1}$Mpc comoving.} 
\end{figure*} 
 
The hierarchical build-up is also thought to determine the morphology of a
galaxy, most noticeably the difference between disk like systems such as spiral
galaxies (some of them barred) and spheroidal systems such as elliptical
galaxies and bulges. This picture envisions that whenever gas is accreted in a
smooth fashion, it settles in rotationally supported disk-like structures in
which gas is slowly transformed into stars. Mergers, however, convert
disks into spheroids. The Hubble type of a galaxy is thus
determined by a continuing sequence of destruction of disks by mergers,
accompanied by the formation of spheroidal systems, followed by the reassembly
of disks due to smooth accretion. This picture of a hierarchical origin of
galaxy morphology has been schematically incorporated in so-called
semi-analytical galaxy formation models used to study the evolution of the
galaxy population, but its validity in a cosmological setting has 
just recently been directly demonstrated (Steinmetz \& Navarro 2002).  
 
Numerical simulations have been an integral part in the detailed analysis of the
virtues of the CDM scenario. Only numerical techniques can account for the highly
irregular structure formation process and for at least some of the complicated
interaction between gravity and other relevant physical processes such as gas
dynamical shocks, star formation and feedback processes. Simulations also
provide the required interface to compare simulations with observational data
and are able to link together different epochs. While simulations of structure
formation on the larger scales have mainly used large massively parallel
supercomputers, studies how individual structures such as galaxies or clusters
of galaxies form in the $\Lambda$CDM scenario have heavily used special purpose
hardware like the GRAPE (=GRAvity PipE) 
family of hardware N-body integrators (Sugimoto \etal 1990).  
 
In this review I will concentrate on some examples, how high resolution gas
dynamical simulation using special purpose hardware have illustrated some
successes but also also some failures of $\Lambda$CDM scenario in reproducing 
structures as observed on the scales of galaxies. 
 
\begin{figure} 
\mbox{\hskip0.7cm\epsfig{file=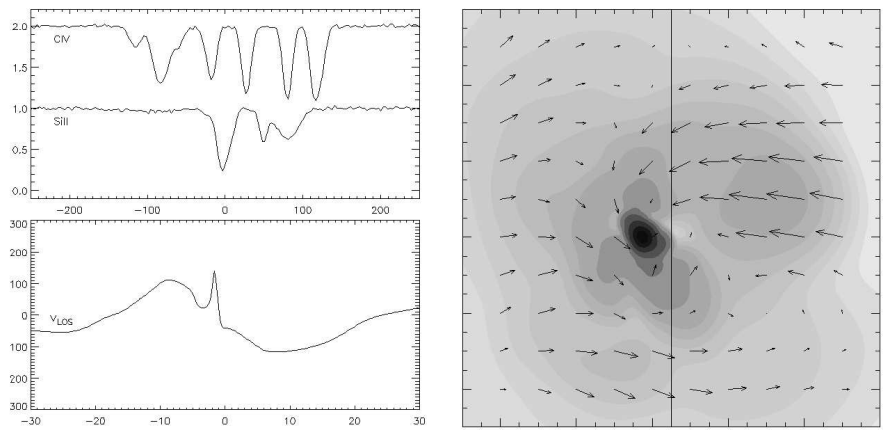,width=12cm}} 
\caption{\label{damped}Right: Color map of the column density distribution in a 
60\,kpc box around a DLAS. Black correspond to \HI\ densities $\log n (\HI) 
> 1.5$), light gray to $\log n (\HI) \approx -3$). Arrows indicate the velocity
field. The solid line corresponds to the line-of-sight (LOS). The lower left
plot shows, the velocity field along the LOS, the upper left plot the absorption
line  in C{\scriptsize IV} 1548 (top) and Si{\scriptsize II} 1808 (bottom). For readability, 
C{\scriptsize IV} has been
displaced  by 0.5 in flux.} 
\end{figure}

\section{Gas dynamical simulations -- The kinematics of damped \Lya\ systems} 
 
Although gas dynamical simulations were considerably successful in explaining
some details of the galaxy formation process, the largest impact so far is
related to the properties of QSO absorption systems. Numerical simulations
can reproduce the basic properties of QSO absorbers covering many orders of
magnitude in column density (Cen \etal 1994; Hernquist
\etal 1996; Zhang, Anninos \& Norman 1995). Indeed, gas dynamical 
simulations were even responsible for a paradigm shift, as QSO absorbers are no
longer considered to be caused by individual gas clouds. Absorbers of different
column density (\Lya forest, metal line systems, Lyman limit systems and damped
\Lya absorption systems) are rather reflecting different aspects of the
large-scale structure of the universe.  While the lowest column density systems
($\log N \approx 12-14$) arises from gas in voids and sheets of the ``cosmic
web'', systems of higher column density are produced by filaments ($\log N
\approx 14-17$) or even by gas that has cooled and collapsed in virialized halos
($\log N > 17$).
 
\begin{figure}[t]
\mbox{\hskip0.7cm\epsfig{ file=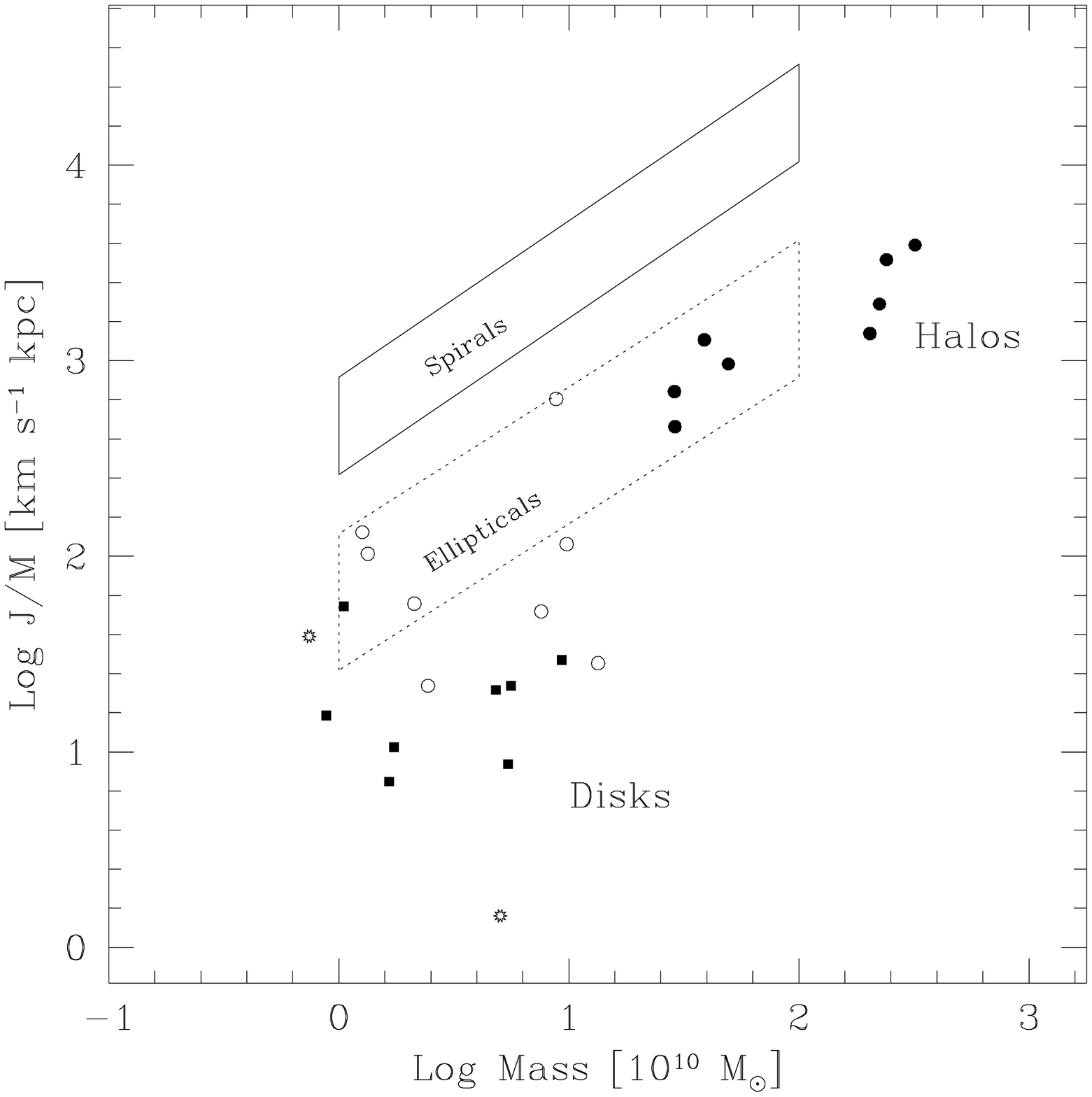,width=5.5cm}}\mbox{\hskip0cm\epsfig{file=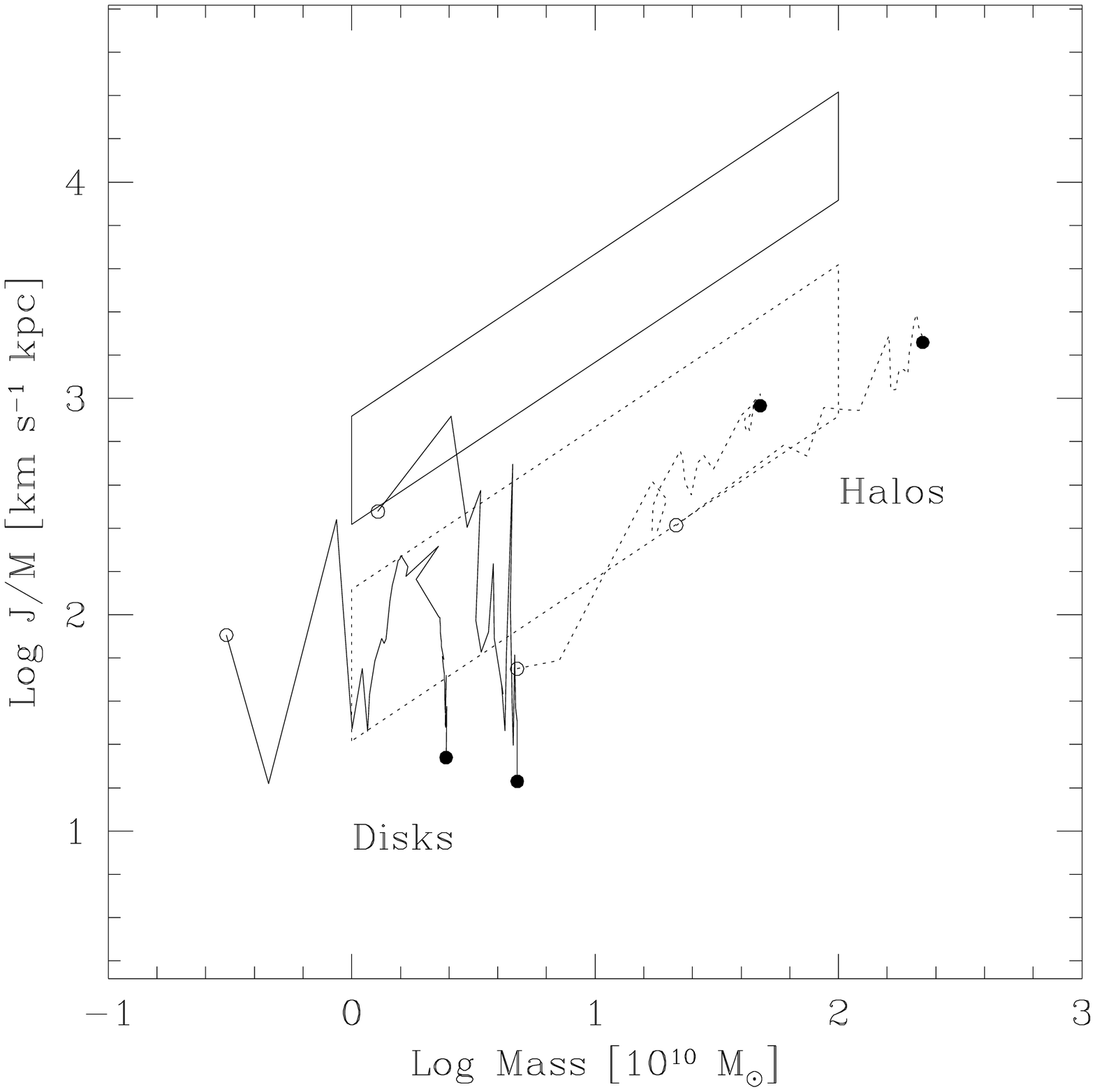,width=5.5cm}} 
\caption{\label{angmom}The specific angular momentum of dark halos and gaseous
disks, as a function of mass. The boxes enclose the region occupied by spiral
and elliptical galaxies, as given by Fall (1983). Open circles, solid squares
and starred symbols correspond to the specific angular momenta of gaseous disks,
solid circles for the hosting dark matter halos.  Right: Evolution of the dark
halo and central gaseous disk in the $J/M$ versus $M$ plane, from $z=5$ (open
circles) to $z=0$ (solid circles).  The mass of the system grows steadily by
mergers, which are accompanied by an increase in the spin of the halo and a
decrease in the spin of the central disk. The latter results from angular
momentum being transferred from the gas to the halo during  mergers.} 
\end{figure} 
 
The kinematics of damped \Lya absorption systems (DLAS) at high redshift serves
as a very nice example to demonstrate how oversimplifying assumptions may lead
to wrong implications on the validity of a cosmological model, while a
full numerical treatment can avoid those artificial contradictions.  DLASs
have often been interpreted as large, high-redshift progenitors of present-day
spirals that have evolved little apart from forming stars (Wolfe
1988). Kauffmann (1996), however, studied the evolution of DLASs in the CDM
structure formation scenario in which disks form by continuous cooling and
accretion of gas within a merging hierarchy of dark matter halos and found that
the total cross section was dominated by disks with comparably low rotation
velocities (typically 70 km/sec).  Prochaska \& Wolfe (1998) however, observed
much larger velocity spreads (up to 200\,km/s) and came to the conclusion that
only models in which the lines-of-sight (LOS) intersects rapidly rotating large
galactic disks can explain both the large velocity spreads and the
characteristic asymmetries of the observed low ionization species (\eg, Si{\scriptsize II})
absorption profiles, in strong contradiction to the prediction of the
(semi-analytical) cold dark matter model.

So how may a numerical simulation solve this problem? The critical assumption
that enters semi-analytical models is that DLASs are equilibrium
disks. Numerical simulations (Haehnelt, Steinmetz \& Rauch 1998) demonstrate
that this is a poor assumption and that asymmetries and non-equilibrium effects
play an important role. Figure 2 shows a typical configuration that gives rise
to a high redshift DLAS with an asymmetric Si{\scriptsize II} absorption profile.  The
velocity width of about 120\,km/s is also quite similar that of typical observed
systems.  No large disk has yet been developed and also the circular
velocity of the collapsed object is only 70\,km/s.  The physical structures that
underlie DLASs are turbulent gas flows and inhomogeneous density structures
related to the merging of two or more clumps, rather than large rotating disks
similar to the Milky Way. Rotational motions of the gas play only a minor role
for these absorption profiles.  A more detailed analysis also demonstrates that
the numerical models easily pass the statistical tests proposed by Prochaska and
Wolfe, \ie, hierarchical clustering, in particular the CDM model, is consistent
with the kinematics of high-$z$ DLASs.
 
\section{Gas dynamical simulations -- The spin of disk galaxies} 
 
While the irregular structure of a galaxy in the process of formation was
helpful to alleviate the problem concerning the kinematics of DLASs, one may
wonder whether it hurts the assembly of large galactic disks at low redshifts as
major mergers are usually associated with the transformation of spirals into
ellipticals (for a review, see Barnes \& Hernquist 1992). 
 
Indeed, gas dynamical simulation are capable of producing disk like structures
(see also section 4), but a closer inspection reveals
that these model disks are too concentrated compared to observed spiral galaxies
(Navarro \& Benz 1991; Navarro, Frenk \& White 1995; Navarro \& Steinmetz
1997). However, this problem is not encountered in semi-analytical models, which
relatively easily can reproduce the sizes of present day galaxies. One may again
ask: What is the difference between the semi-analytical model (Mo, Mao \& White
1997) and the numerical simulations, in particular since they are based on the
same structure formation model? Again, the reason can be found in the
assumptions that enter the semi-analytical models: Semi-analytical models assume
that gas collapses under conservation of angular momentum (Fall \& Efstathiou
1980), an assumption that, as will be shown, is only very poorly fulfilled.
 
Figure 3 (left) shows the specific angular momentum of dark matter halos and of
their central gaseous disks at $z=0$, as a function of mass.  If, as suggested
by Fall \& Efstathiou (1980), the collapse of gas would proceed under
conservation of angular momentum, the baryonic component would have the same
specific angular momentum $J/M$ as the dark matter, however, its corresponding
mass would be a factor of 20 smaller (for $\Omega_{\rm bary} = 0.05$,
$\Omega_0=1$\footnote{The angular momentum problem has first been demonstrated
in simulations based on an $\Omega_0=1$ CDM scenario. The problem is, however,
generic to all hierarchical models and can also be seen in more recent simulations
of galaxy formation in a $\Lambda$CDM cosmology}). 
These disks would be located only slightly below the box for
spiral galaxies. However, Figure 3 demonstrates clearly that the spins of
gaseous disks are about an order of magnitude lower than that. This is a direct
consequence of the formation process of the disks. Most of the disk mass is
assembled through mergers between systems whose own gas component had previously
collapsed to form centrally concentrated disks. During these mergers, and
because of the spatial segregation between gas and dark matter, the gas
component owing to dynamical friction 
transfers most of their orbital angular momentum to the surrounding
halos. While the specific angular momentum of dark matter halos increases with
decreasing redshift, that of gaseous disk decreases (Figure 3, right).

Up to now, it is unclear how this discrepancy can be overcome. One potential
solution envisions the presence of strong feedback processes (Weil \etal, 1998,
Thacker \& Couchman 2002) that prevent gas form clumping early on and that thus
reduce the transfer of angular momentum due to dynamical
friction. Alternatively, the low angular momentum gas may be ejected during
early phases of galaxy formation or absorbed in the bulge component
(Dominguez-Tenreiro \etal 1998, van den Bosch \etal, 2002). If these solutions
turn out to be unable to solve this problem, one may also have to consider
modifying the CDM paradigm on the smallest scale, i.e., reducing number of
mergers, \eg, by considering {\sl warm dark matter} (WDM; see, e.g.,
Somer-Larson \& Dolgov, 2001).
 
\section{Gas dynamical simulations including star formation -- The origin of galaxy morphologies} 
 
\begin{figure*} 
\begin{center} 
\psfig{file=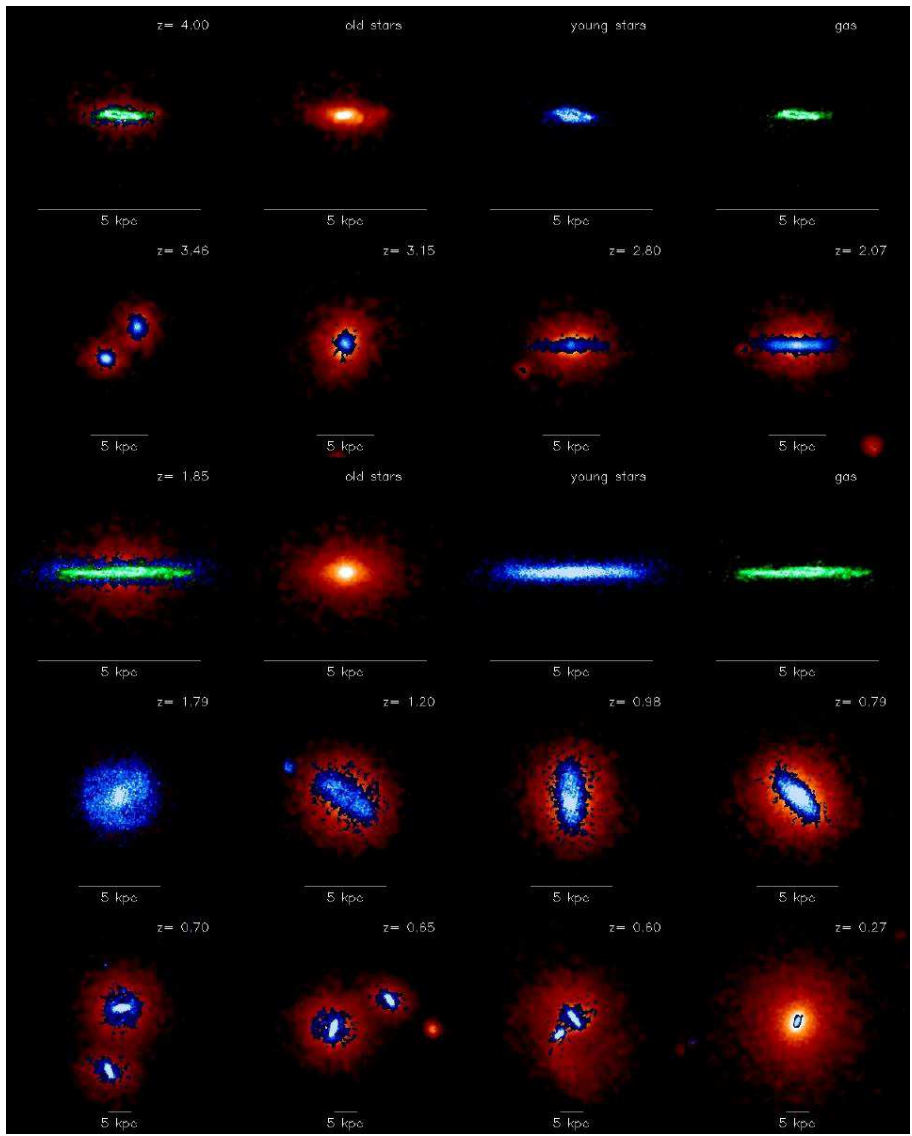,height=15cm} 
\end{center} 
\caption{\label{number} Surface density of the gaseous and stellar components of
$\Lambda$CDM   halo at various epochs. Horizontal bars in each panel are 5
(physical) kpc long and indicate the scale of each figure. Rows 2, 4 and 5 show
time sequences near some key evolutionary stage. Rows 1 and 3 decompose a galaxy
at a particular redshift (left) into its constituents: old stars, young stars
and gas (from left to right). Top row: The most massive progenitor at z=4 shown
edge-on. Second row: The formation of a bulge and the rebirth of a disk.  Third
row: The appearance of the galaxy at z=1.8, seen edge-on.  Fourth row: The tidal
triggering of bar instability by a satellite resulting in the emergence of a
rapidly rotating bar. Bottom row. A major merger and the formation of an
elliptical  galaxy.} 
\end{figure*}

A more realistic modeling of galaxy formation requires, of course, the inclusion 
of star formation and related feedback processes such as supernovae or stellar
winds. Unfortunately, the physics of star formation and of the interaction of evolving stars
with the interstellar medium is only ill understood. Explicit inclusion of these effects in
direct numerical simulations can only be accomplished in the form of
numerical ``recipes'' that contain a number of free parameters which can be
calibrated e.g. such that Kennicutt's (1998) relation between the H{\scriptsize I}
surface density and the star formation rate per unit area (Schmidt's law) is 
reproduced in isolated galaxy test cases.

The level of detail with which it is now possible to model the formation of individual
galaxies is  illustrated in Figure 4 and 5 which convincingly demonstrate that 
numerical simulations are now able to capture the main stages of
galaxy formation, and to resolve their main morphological components. 
In particular, the following phases can be identified: 
 
\begin{itemize} 
\item {\bf Formation of first disks.} At $z\approx 4$ the first proto galactic
clumps are  numerically resolved (i.e.~more than 500 particles per halo). 
Small disks (diameter $\approx 3\,$kpc) form near the centers of these clumps. 
 
\item {\bf Mergers and bulge formation.} The proto galactic clumps frequently
merge ($z\approx 3$)  with objects of comparable size and mass. The small disks
at the  center are destroyed in this process and a central spheroidal component 
forms. 
 
\item {\bf Regeneration of disks.} In the absence of major mergers, disk-like
structures  around the bulge regenerate from gas accreted both smoothly and
through minor mergers ($z\approx 2$). These objects feature all the major
dynamical components of a bright spiral  galaxy like the Milky Way: a
rotationally  supported disk of young stars, a centrally concentrated bulge, and
a stellar halo of older stars. 
 
\item {\bf Minor mergers and the formation of rapidly rotating bars.} Tidal
torques during the close encounter with a minor satellite (mass ratio $\approx
1:10$) can trigger the formation of a rapidly rotating bar in the disk. The bar
pattern persists long after the disruption of the culpable satellite. The bar
extends out to about 2.5 kpc and has a corotation radius of slightly less than 3
kpc, implying, in agreement with the few barred galaxies where this ratio has
been measured (Debattista \& Sellwood 1999), that the bar is ``fast". The bar
pattern is clearly visible for more than 30 orbital periods, a consequence of
repeated triggering by orbiting satellites aided by the fact that the baryonic
component dominates the central potential (75\% of the mass within 3 kpc is in
the disk). Although the process described here is likely not the only bar
formation mechanism acting in real galaxies, the simulation shows nevertheless a
direct link between the presence of a bar pattern and the existence of
satellites that may act as tidal triggers.
 
\item {\bf Major mergers and the formation of elliptical galaxies.} Mergers
between close to equal mass galaxies effectively destroy disks and give rise to
the formation of an elliptical galaxy. Residual gas in the disk progenitors is
effectively channeled to the center where it is transformed to stars in a
starburst like manner ($z=0.25$). The resulting object is mainly composed of
``old'' stars with a sprinkle of young stars at the center that formed during
the last major merger.
 
\end{itemize} 

Figure 4 illustrated the formation history for a galaxy that encountered a major
merger at relatively low redshift ($z\approx 0.7$) and experience little accretion
of mass afterwards. The morphology at $z=0$ thus corresponds to that of an
elliptical galaxy. Figure 5 shows a simulation of a different halo that did not
experience any major merger after redshift 1.5, but more mass has been added by smooth
accretion between $z=1.5$ and $z=0$. Consequently, its final morphology is
similar to that of a major disk galaxy.

The examples shown in Figures 4 and 5 demonstrate that high resolution
simulations of the formation of individual galaxies are now capable to 
directly validate the long standing hypothesis, that depending on their merging
history galaxies may repeatedly change their morphology and that morphology may
be a transient phenomenon.  These simulations also highlight some of the
potential shortcomings and open issues of hierarchical galaxy formation
scenarios. For example, is the large fraction of stars observed in disks today
consistent with the `lumpy' accretion histories expected in CDM-like scenarios?
Can one account for the observed frequency, size, and dynamical properties of
`pure disk' (bulge-less) galaxies?  Can variations in accretion history with
environment account for the morphology-density relation?  These questions are
critical to the success of the hierarchical model and remain challenging puzzles
to be elucidated within this so far highly successful paradigm. It will take a
substantive computational effort to address them through direct simulation, but
one that is within reach of today's technological capabilities.

\begin{figure} 
\mbox{\epsfig{file=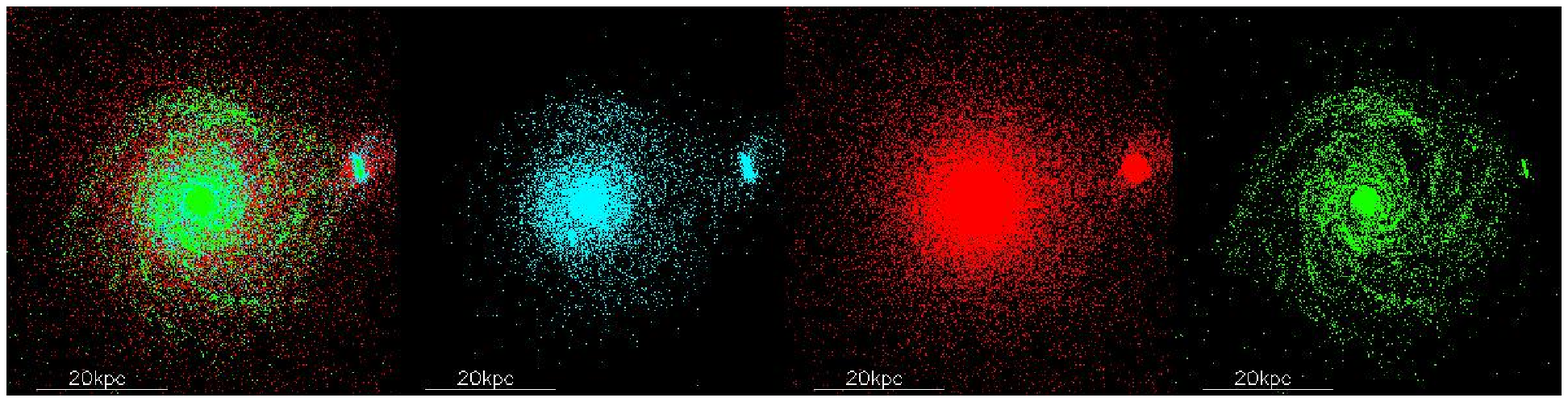,height=3.3cm}} 
\vskip 0.1cm
\mbox{\epsfig{file=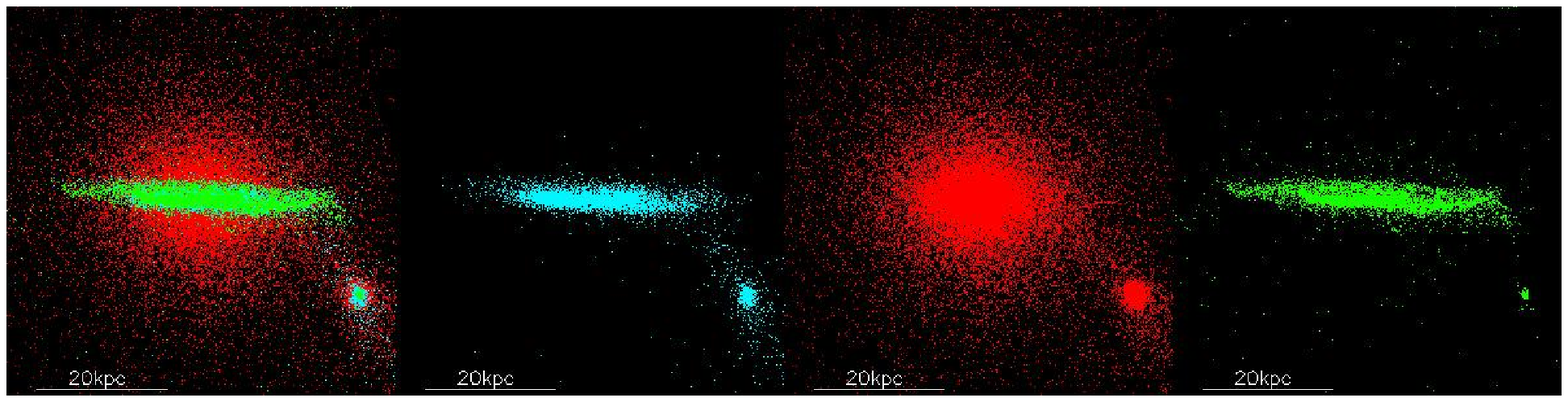,height=3.3cm}} 
\caption[\label{xy}]{Face on (top) and edge-on (bottom) view of a simulated disk galaxy at 
redshift zero. Left: distribution of all stars and gas particles; Middle-left:
distribution of stars younger than 0.6 Gyr; Middle-right: distribution of stars
older than 5 Gyr; Right: distribution of gas particles}
\end{figure}

\section{Gas dynamical simulations including star formation -- The origin 
of disk galaxy scaling laws} 
 
The success of such a model can be further assessed by testing, to what extent
such a model can reproduce scaling relations that link total luminosity,
rotation speed, and angular momentum of disk galaxies such as the Tully--Fisher
(TF) relation.  Figure 6 shows the results of such an investigation, the
simulated $I$-band TF relation at $z=0$ for the $\Lambda$CDM and for the
$\Lambda$WDM scenario. The
simulated TF relation is compared with the data of Giovanelli \etal~(1997),
Mathewson, Ford \& Buchhorn (1992) and Han \& Mould (1992).  The slope and
scatter of the simulated TF relation are in fairly good agreement with the
observational data.  This result also holds in other bandpasses: the model TF
relation becomes shallower (and the scatter increases) towards the blue, just as
in observational samples (see Steinmetz \& Navarro 1999).  The model TF
relations are also very tight. In the I-band the {\sl rms} scatter is only 0.25
mag, even smaller than the observed scatter of $\sim 0.4\,$mag. This must be so
if the results are to agree with observations: scatter in the models reflects
the intrinsic dispersion in the TF relation, whereas the observed scatter
includes contributions from both observational errors and intrinsic
dispersion. If, as it is usually argued, both effects contribute about equally
to the observed dispersion in the TF relation, then the intrinsic scatter in the
I-band should be comparable to the $\sim 0.25\,$mag found in the models.
 
In addition, the zero-point is in rough agreement with observational data, at
closer inspection the simulated galaxies appear to be $\sim 0.3\,$mag too
dim. This disagreement can be accounted to a still sub optimal modeling of the
star formation history in these models. Due to the particular parameterization
of star formation, a large fraction of stars form already at high redshift
resulting in a fairly high stellar mass-to-light ratio of 2-2.5 in
I-band while observations indicate a value closer to 1-1.5 (Bottema 1997). It is also
interesting that the zero-point of the TF relation seems to strongly depend on the assumed
cosmological scenario. A similar simulation performed for a $\Omega=1$,
$\Lambda=0$ CDM scenario results in a very in serious disagreement of the zero
point with the data, the simulated $\Omega=1$ TF relation being almost two
magnitudes too faint at given rotation speed.
 
Another interesting clue on the origin of the Tully-Fisher relation is that the
small scatter and the zero-point (for a given cosmological scenario) is only
weakly dependent on the particular star formation and feedback parameterization:
Very similar results can be obtained for quite different star formation
prescriptions (Navarro \& Steinmetz, 2000).  This is a surprising result since
the fraction of baryons that ends up in a disk strongly depends on the
particular feedback model and can vary by factor 3-5 between different
parameterizations.
 
The weak dependence of the scatter on the feedback parameterization is an
immediate consequence of how the rotation velocity of the galactic disk responds
to changes in the ratio between stellar and dark mass. As more and more baryon
assemble in the central stellar disk, their luminosity increases but also the
disk rotation velocity due to the gravity of the additional matter near the
center. This effect is amplified as the additional gravity also pulls dark
matter towards the center. Consequently a variation in the star to dark matter
fraction results in a shift predominately parallel to the TF relation which
thus does not cause substantial additional scatter (see Navarro \& Steinmetz
2000 for a detailed discussion)
 
\begin{figure} 
\mbox{\hskip3cm\epsfig{file=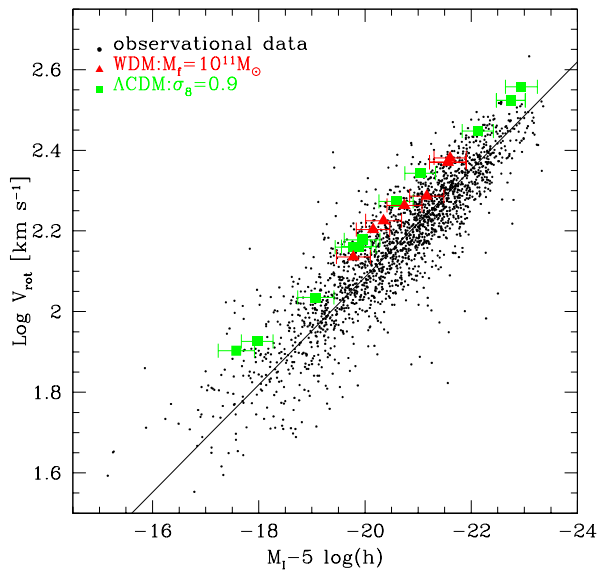,height=8cm}} 
\caption[\label{tully}]{I-band TF relation at $z=0$ for a $\Lambda$CDM and a
$\Lambda$WDM scenario. The error bars in the simulated data span the difference
in magnitudes that results from adopting a Salpeter or a Scalo  IMF.}
\end{figure} 
 
\section{Summary and Conclusions} 
 
I presented some results of recent efforts to model the formation and evolution of
galaxies in a hierarchical structure universe using high-resolution computer
simulations. I demonstrated that only numerical simulation can take full account
of the dynamics of the formation process and the complicated interplay between
different physical processes such as, \eg, accretion and merging, star formation
and feedback, photo heating and radiative cooling. Observational data can easily
be misinterpreted if these effects are not properly included.  For example, the
apparent inconsistency of hierarchical structure formation models and the kinematics of
high-$z$ damped \Lya absorption systems could be easily solved by properly
accounting for the complicated non-equilibrium dynamics of galaxies in the
process of formation. Interestingly, the opposite is the case for the example
of the galaxy spin. The simplifying assumption of collapse under conservation of 
angular momentum combined with success of this model in explaining the sizes of disk 
galaxies may lure someone to consider the origin of the
sizes of disk galaxies as being solved. However, at closer inspection using numerical 
simulation, the physics behind the sizes of disk galaxies may be far more
complicated. Maintaining the hierarchical build-up of galaxies and
simultaneously avoiding substantial exchange of angular momentum from the gas to
the dark matter due to mergers appears to be a major challenge to the scenario,
and it cannot be excluded that the angular momentum crisis may finally
lead to substantial revisions of the hierarchical structure formation scenario. 

\begin{acknowledgements}  
 
This article includes work from collaborations with V.~Eke, M.~Haehnelt, 
J.~Navarro and M.~Rauch.  This work has been supported by the National
Aeronautics and Space Administration under NASA grant NAG 5-7151 and NAG
5-10827, by the National Science Foundation under NSF grant 9807151, and by
fellowships from the Alfred P.~Sloan Foundation and the David and Lucile Packard 
foundation. 
\end{acknowledgements}

\end{document}